\documentclass [12pt]{article}    

\usepackage{graphicx}

\graphicspath{ {./images/} }

\begin{document}

\title{Rotation curves of galaxies  via Reissner-Nordstrom induced gravity and an alternative explanation of Dark Matter}
\maketitle

\centerline {\author{\bf J. Buitrago}}

%`\author{J. Buitrago}
%\address{University of La Laguna, Faculty of Physics. 38205, La Laguna (Tenerife) Spain, (jgb@iac.es)}
%\footnote{jgb@iac.es}

\bigskip

\noindent{University of La Laguna, Faculty of Physics. 38205, La Laguna (Tenerife) Spain. 

\bigskip

jgb@iac.es

%\maketitle

%\address{jgb@iac.es}
%\address{$^2$Affiliation2 details for second author to appear here, Cairo 11566, Egypt}
%\address{$^3$Affiliation3 details for third author to appear here, P.O.~Box 43, Egypt}
%\address{$^4$Affiliation4 details  for fourth author to appear here, California 92521, USA}

\begin{abstract}
The dynamics of a neutral test particle in the spacetime geometry  to a central massive and charged object (Reissner-Nordstrom Metric) is examined. For a radial distance $r = Q^2/{M}$ (in natural units) the gravitational force is null, independently of the value of $G$, and repulsive below this value. It is shown that within typical atomic and molecular distances, there is a repulsive force albeit negligible in comparison with the electromagnetic one ruling the atomic world. For an eventual extremal black hole having a mass equal to the Planck Mass a limit to electric charge equal to $1(MeV)^0$ is found.
At the galactic scale and for galaxies with a compact central nucleus with mass below or of the order of ${M_\odot}$, the repulsive force can reproduce the flat rotation curve of stellar orbits observed in many galaxies. %This is place reserved for Abstract. This samle document describes some general features of \LaTeX, at the same time showing how LHEP template output format when standrad \LaTeX command is used. LHEP cls tried to output section, subsection, figure and table caption to appear in attractive manner using fonts that fits to our requirements.
\end{abstract}

%\maketitle

%\begin{keyword}
%General Relativity \sep Dark Matter \sep Galactic Rotation Curves
%\doi{10.2018/LHEP000001}
%\end{keyword}

\section{INTRODUCTION}

The main object of this work is to present some new consequences of a repulsive gravitational force induced by the Reissner-Nordstrom spacetime which had been reported some 11 years ago \cite{daniela}, apparently not receiving the  attention by the scientific community that such a result deserved. 
The more important consequence of the mentioned work is that 
radial geodesics followed by test neutral particles in the spacetime induced by the Reissner Nordstrom metric (RN) have a repulsive behavior below
 a distance to the center of $Q^2/M$ (in natural units),  being $Q$ and $M$ the charge and mass of the central object respectively, the force being null for this distance.
 
 In this article new features of this sort of repulsive gravity and some of its consequences for physics and astrophysics will be explored.  A non phenomenological alternative to the so called MOND theory \cite{milgrom1} will be presented and also call the attention about the possibility of primordial collapsed objects with a net electrical charge (together with some other possibilities, see the FINAL COMMENTS at the end) that perhaps in the early universe might have been significative.
 
 Quite recently, a model (Cappelluti et al. 2022) \cite{cappelluti} has been suggested in which primordial black holes with a broad birth mass function ranging in mass from a fraction of a solar mass to about $10^6 M_{\odot}$, peaking near the Chandrasekhar mass of $1.4 {M_\odot}$, constitute the Dark Matter component in the Universe. On the other hand, for some years the so called MOND theory (Milgrom 1983) \cite{milgrom1} has offered paradigmatic albeit phenomenological explanation of the peculiar flat rotation curves  of galaxies.

One of the main results of this study is that it is not possible to fit the flat galactic rotation curves beyond 1-2 Solar masses of the collapsed objects and, surprisingly, we also find that the condition  $GM^2<Q^2$  corresponding to Naked Singularities instead of black holes must hold thus pointing to a different mechanism for the formation of supermassive black holes in galaxies.

Since its invention, the MOND phenomenological theory aiming to an explanation of the peculiar rotation curves in many galaxies have been the subject of many additional studies and controversies which are out of the scope of this work. What we shall present in this study, is from the theoretical standpoint self-consistent its validity depending only on whether RN collapsed objects exist in the universe.

%\begin{itemize}
%\item {\bf What is \LaTeX\,?}
%\item {\bf Basic usage and syntax}
%\item {\bf Modes and environments}
%\item {\bf Newcommands}
%\item {\bf Cross-referencing}
%\item {\bf Packages}
%\item {\bf Importing graphics}
%\item {\bf Tables and figures}
%\item {\bf Pictures}
%\item {\bf Where to learn more}
%\end{itemize}

\section{RADIAL GEODESICS}%{{ WHAT IS \LaTeX\,?}}
%\subsection*{What \LaTeX\, is NOT:} 

In spherical coordinates $x^\alpha=(t,r,\theta,\phi)$, (natural units $c=\hbar=1$ will be used), the element of arch parametrized by proper time $\tau$ is

\begin{equation}
d\tau^2 =\left(1-\frac{2GM}{r}+\frac{GQ^2}{r^2}\right )dt^2-\left(1-\frac{2GM}{r}+\frac{GQ^2}{r^2}\right )^{-1}dr^2 - \ r^2d\theta^2-r^2\sin^2\theta  d\phi^2.
\end{equation}
(In natural units, the gravitational constant is $G=6.76088\times 10^{-45} MeV^{-2}$), radial distances are given in $MeV^{-1}$. 1 kpc = $1.563\times 10^{20}MeV^{-1}$).

For the study of radial and circular geodesics, a convenient start point is the Lagrangian \cite{chandra}

\begin{equation}
{\cal L} =\frac{1}{2}\left[\left(1-\frac{2GM}{r}+\frac{GQ^2}{r^2}\right )\dot t^2-\left(1-\frac{2GM}{r}+\frac{GQ^2}{r^2}\right )^{-1}\dot r^2 -r^2\dot\theta^2-r^2\sin^2\theta \dot \phi^2
\right]
\end{equation}
(dot meaning derivative respect to proper time $\tau$).

Together with the Euler-Lagrange equations
\begin{equation}
\frac{d}{d\tau} { \frac{	\partial {\cal L}}{\partial \dot x^\alpha}-\frac{\cal \partial L}{\partial x^ \alpha} = 0}.
\end{equation}

For $x^0 = t $, if we restrict to radial geodesics and since the coordinate time does not explicitly appears in the lagrangian, the corresponding canonical momentum
is a constant of the motion $\widetilde E$ that can be interpreted as the energy 
per unit mass:
\begin{equation}
 u_0 = \frac{\partial \cal L}{\partial \dot t} = \left(1-\frac{2GM}{r}+\frac{GQ^2}{r^2}\right )u^0 = \widetilde E.
\end{equation}
If we drop the test particle from infinity on a radial trajectory beginning
 at rest, the constant $\widetilde E$ is simply unit. Then, from the last equation and equation (1):
 \begin{equation}
 u^0 = \frac{dt}{d\tau}	= \frac{1}{ \left(1-\frac{2GM}{r}+\frac{GQ^2}{r^2}\right )}
 \end{equation}
 and
 \begin{equation}
 u^1=\frac{dr}{d\tau} = \sqrt{\frac{2GM}{r}-\frac{GQ^2}{r^2}}.
 \end{equation}
 The non null Christoffel symbols for the radial coordinate are:

\begin{equation}
	\Gamma^0_{01}=\frac{2GMr-2GQ^2}{2r(r^2-2GMr+GQ^2)}, 
\end{equation}
$$
\Gamma^1_{00}=\frac{1}{2}(1-\frac{2GM}{r}+\frac{GQ^2}{r^2}).(\frac{2GM}{r^2}-\frac{2GQ^2}{r^3})
$$

$$
\Gamma^1_{11}=-\frac{GMr-GQ^2}{r(r^2-2GMr+GQ^2)}.
$$

From the precedent relations and the corresponding geodesic equation, the radial acceleration relative 
  to the proper time is
  \begin{equation}
 \frac{d^2 r}{d\tau^2} = \frac{GM}{r^2} - \frac{GQ^2}{r^3}.
\end{equation}
Irrespective of the value of the gravitational constant $G$, the right hand side is null for a radial distance

\begin{equation}
r_{NF} = \frac{Q^2}{M}.
\end{equation}
For radial distances below this value no circular orbits can exist. For $r$ slightly above this value, the repulsive term produced by the charge (of either sign) makes the orbital velocity much smaller than the  newtonian or Schwarzschild orbit. As the repulsive effect decreases faster with distance, far from the source, orbits are newtonian. 
\section{CIRCULAR ORBITS}

As obtained in \cite{daniela}, for stable circular orbits, the orbital velocity is given by
\begin{equation}
v = \sqrt{\frac{GMr-GQ^2}{r^2-2GMr+GQ^2}}
\end{equation}
For $r < r_{NF}$ no circular orbits exist. On the other hand, it is well known that the quadratic expression in the lower part of the precedent equation determines the Event and Cauchy horizons of the RN spacetime. Solving the quadratic equation for $r$:
\begin{equation} \label{cauchy}
	r = GM\pm {\sqrt{G(GM^2-Q^2)}}
\end{equation}
For positive values of  $GM^2-Q^2$, both kind of horizons exist. The case $GM^2=Q^2$ is the so called Extreme Black Hole while for $GM^2<Q^2$ we have a naked singularity.
For an analysis of the three physical situations just mentioned, it is convenient to rewrite the circular velocity equation in the alternative form:
\begin{equation}
	v=\sqrt{{\frac{\frac{GM}{r}\left(1-\frac{Q^2}{M}\frac{1}{r}\right)}{1-\frac{GM}{r}\left(2-\frac{Q^2}{M}\frac{1}{r}\right)}}}.
\end{equation}
We introduce next a dimensionless parameter $x\ge 1$ so that
\begin{equation}
	r=x\frac{Q^2}{M},
\end{equation}
thus expressing the orbital velocity as

\begin{equation}
	v=\sqrt{{\frac{\frac{GM^2}{xQ^2}\left(1-\frac{1}{x}\right)}{1-\frac{GM^2}{xQ^2}\left(2-\frac{1}{x}\right)}}}.
\end{equation}
%
%For an extremal black hole, from (\ref{cauchy}), $Q=\sqrt{G}M$ and there is only one horizon located in $r=GM$. 
%

The previous results are valid for any physical system where mass and charge are present (As commented below we are at the present cosmic time far from the Planck scale where quantum effects begin to dominate). For instance, at the atomic scale, $r_{NF}$ for the Hydrogen atom would be located at  0.000153 cm. This is a rather curious result revealing that in the atomic and molecular world, electrons live under repulsive gravitational forces (possibly quite negligible). To further clarify this point, for $r=0.52\times 10^{-8} cm$ (Bohr Radius),
 the classical Coulomb force is
 \begin{equation}
 \frac {e^2}{r_B^2}=-1.014\times 10^{17} (MeV)^2,
 \end{equation}
while the repulsive gravitational one (multiplying by the electron mass in equation (8)) is $1.373 \times 10^{-18} (MeV)^2$. From this simple-minded result, it seems that the repulsive component is probably outside any attempt of experimental verification.

At the atomic level the zone of null force lies well outside the atomic region, and consequently, all atoms are immersed in the repulsive zone. Furthermore, a simple calculation in equation (\ref{cauchy}) reveals that from the RN metric standpoint both protons and electrons are naked singularities avoiding Penrose's Cosmic Censorship Hypothesis which is assumed to be only applicable in the macroscopic domain.

If we ask the question of what should be the minimum mass and charge of a stable collapsed object such that it would be an extreme RN black hole, a natural choice seems the Planck Mass, from equation (\ref{cauchy}); the condition is
\begin{equation}
	(GM)^2-GQ^2=0,
\end{equation} 
and then 
\begin{equation}
	Q=\sqrt{G}M,
\end{equation}
generally valid for any RN extreme black hole. For $M$ being equal to the Planck Mass, $M_p=1/ \sqrt G$, and we find 
\begin{equation}
Q=1 (MeV)^0
\end{equation}
(Note that $r_{NF}$ is also located at the only horizon at the Planck Length $L_p=1/{M_p}$).

As it is only $G$ which determines the value of $M_p$, the previous numerical value is independent of any specific value of the gravitational constant, thus setting a lower limit for the electric charge of an extreme RN collapsed object. Recalling that the electron charge $e=0.08542 (eV)^0$, $Q=1$ is equal to 
$11.706\times 10^6$ times the electron charge. Note that in Planck Units electric charge is absent as Planck himself considered mass and charge inconmensurable quantities. %Since charge is a Lorentz invariant quantity $Q=1$ can be interpreted as the maximum electric charge that an elementary particle can hold in the same sense that $c=1$ is the maximum velocity% 
 As we cannot go beyond the Planck scale of energy about $10^{19} GeV$ where quantum effects begin to dominate, this is perhaps a sensible result for the initial conditions in the Universe.
 
 For any extremal RN black hole, in terms of the dimensionless coordinate $x$, the velocity curve is always the same irrespective of the mass (see fig 1):
\begin{equation}
	v= \sqrt{\frac{\frac{1}{x}\left(1-\frac{1}{x}\right)}{1-\frac{1}{x}\left(2-\frac{1}{x}\right)}},
\end{equation}

\begin{figure}[h]
  \includegraphics [width=11cm]{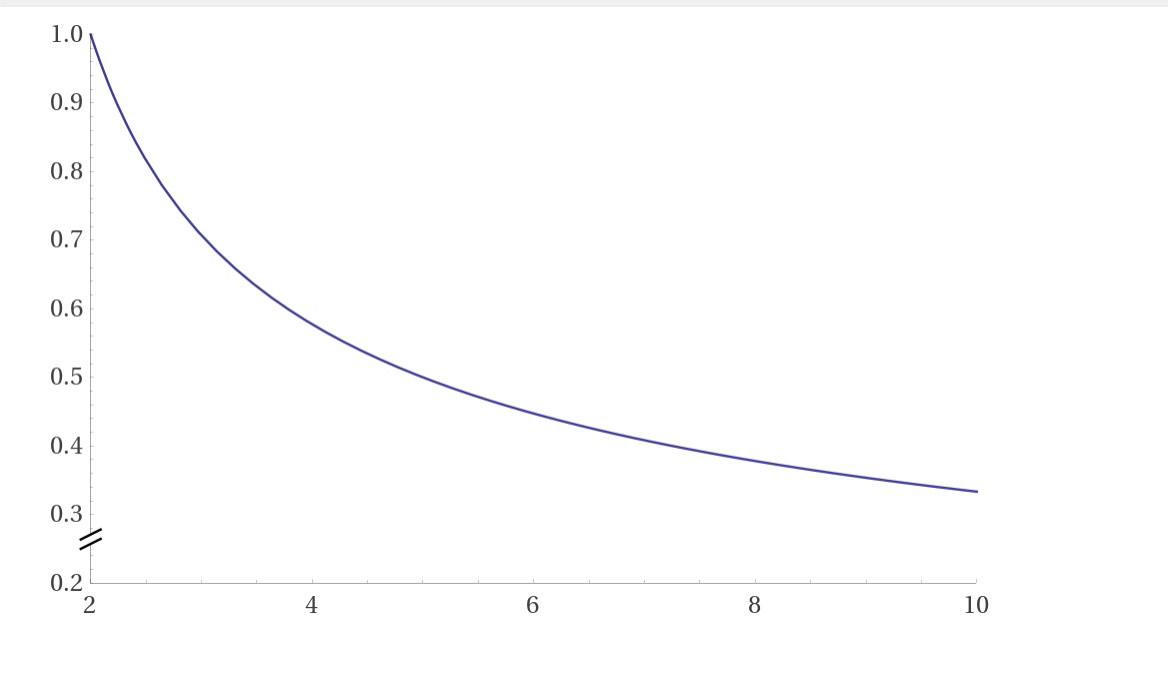}
  \centering
  \caption{Universal rotation velocity for Extreme Black Holes. Label of Horizontal axis is the dimensionless parameter $x$. Vertical axis: Velocity as a fraction of the velocity of light. (For $x=2$, $v=1$)}.
\end{figure}

The question now arises as whether circular orbits could exist in the presence of a naked singularity (the very existence of naked singularities is forbidden by the questionable cosmic censorship hypothesis). To clarify such an intriguing possibility, we obtain the derivative respect to $r$ of equation (10) and solve for $r$ the resulting expression equated  to zero, thus obtaining the  radial distance corresponding to the maximum and minimal orbital velocity:
\begin{equation}
r=\frac{1}{M}Q\left(Q\pm \sqrt{Q^2-GM^2}\right).
\end{equation}
%
%(La solucion se ha comprobado con Derive y Wolfram)
%{\bf Es lo primero que hay que comprobar con las gráficas si la posición del máximo es la prevista }

%

Since no circular orbits exist below $r_{NF}$, the maximum velocity correspond to the $+$ sign solution. We note further that the reality condition $Q^2 \geq GM^2$ in the last equation, opposite to the same condition in equation (11), exclude the possibility of two horizons. If we are going to have some interval in $r$ of circular orbits, only the extreme case of a naked singularity is allowed.

\section{FIT TO OBSERVATIONAL RESULTS}

To see whether what have been exposed until now might have something to do with observational reality at the galactic scale we shall consider two illustrative cases which were used to manifest the adequacy of MOND (Milgrom and Sanders 2007) \cite{milgrom2} for explaining the experimental shape of the velocity rotation curve of some low mass galaxies.

\begin{figure}[h]
  \includegraphics [width=11cm]{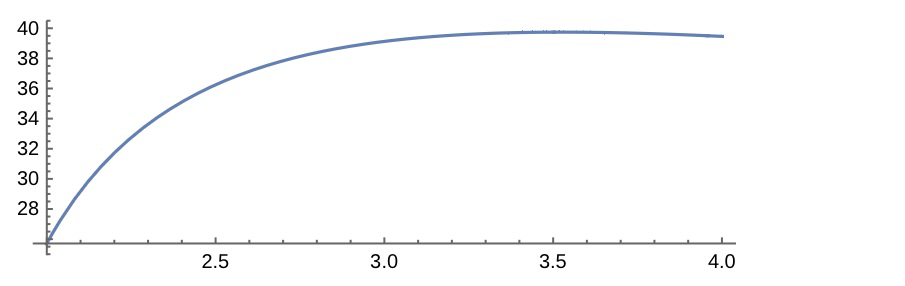}
  \centering
  \caption{Rotation curve fitting the MOND and experimental results  of NGC 3741 and KK98 250. Horizontal axis distance in kpc and Vertical axis velocity in km/s}
\end{figure}

The first case to consider is the extrapolation of MOND to very low-mass spiral galaxies as explained in Milgrom and Sanders (2007)\cite{milgrom2}. They test MOND from the experimental data of four spiral galaxies with a baryonic mass below $4{M^8_\odot}$. We have taken two of them KK98 250 and NGC 3741 (no reference found to the existence of a black hole in any of them) both with rotation curves extending from 0 to 4 kpc (see Fig.1 in \cite{milgrom2}). To fit the rotation curves, I considered an RN collapsed object with a 0.00258 $M_\odot$ mass and a total charge equivalent to the charge of $1.04\times 10^{40}$ electrons (see Fig.2).
The radius of null force and null velocity is $Q^2/{M}=1.56$ kpc. From the mass and charge values fitting the experimental curves, it is readily seen that they correspond to a Naked Singularity.

\begin{figure}[h]
  \includegraphics [width=11cm]{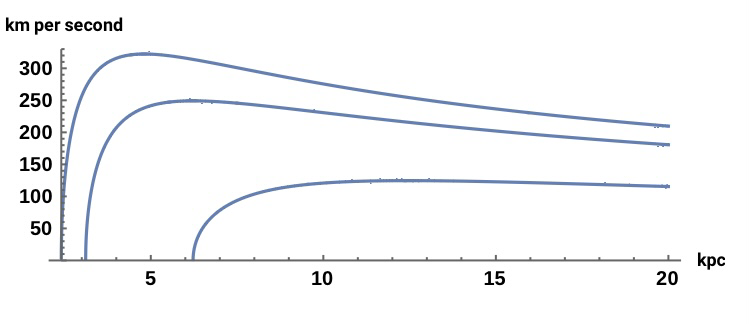}
  \centering
 \caption{Rotation curve fitting some of the data relative to spiral galaxies  (see Figure 4 in \cite{ann}). $M_1=2.587\times 10^{59} MeV$, $M_2=2.00\times10^{59} MeV$ , $M_3=1.00\times 10^{59} MeV$, corresponding to 0.232, 0.179 and 0.090 Solar Masses, respectively. $Q=9.8536\times 10^{39} MeV^0$, $M_{\odot}= 1.11542\times 10^{60} MeV$}
\end{figure}

\begin{figure}[h]
  \includegraphics [width=11cm]{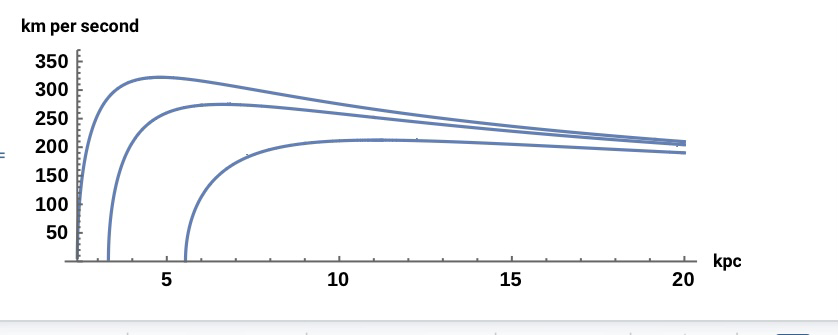}
  \centering
 \caption{Rotation curves (charge increasing from the upper to the lower part), $Q_1=9.8536\times 10^{39} MeV^0$, $Q_2=11.5536\times 10^{39} MeV^0$, $Q_3=14.9536\times 10^{39}  MeV^0$}
\end{figure}

\begin{figure}[h]
  \includegraphics [width=11cm]{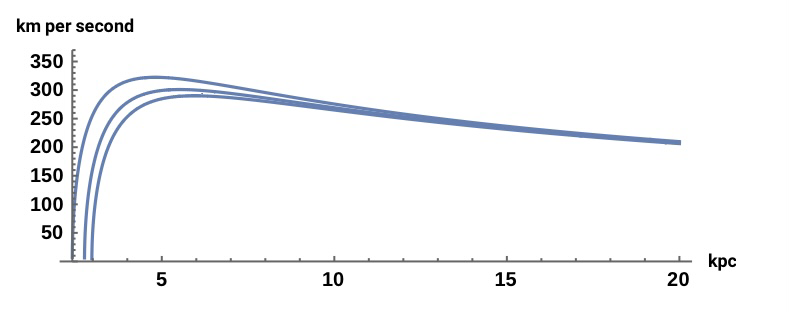}
  \centering
 \caption{Rotation curves (charge decreasing from the upper to the lower part), $Q_1=9.8536\times 10^{39} MeV^0$, $Q_2=9.05363\times 10^{39} MeV^0$, $Q_3=8.55363\times 10^{39}  MeV^0$}
\end{figure}

The other three figures are intended to describe the relative influence of  mass and charge in shaping the rotation curves (For real observational curves see for instance \cite{ann}, figure 4 and \cite{sofue}). The upper curve is always the same in the three figures corresponding to a collapsed central object of 0.232 Solar Masses. In figure 3, we keep the charge constant and decrease the mass in the middle and lower curves. In figures 4 and 5 we keep the 0.232 $M_{\odot}$ mass constant while increasing and decreasing the charge, respectively. Note in figure 4 how the increase in charge flattens the rotation curves (the decrease in charge makes the curves in figure 5 close together).

In figures 3,4, and 5, the radial distance of null force $r_{NF}=Q^2/M$ in the upper curve is located at $3.753\times 10^{20}MeV^{-1}$, equivalent to $2.401$ kpc (1 kpc =$1.563\times 10^{20}MeV^{-1}$). 

A simple look at equation (14) reveals that all rotation curves having the same value of the dimensionless quantity $GM^2/Q^2$ will have the same shape. However, the radial distance to the source can be quite different. For instance, if we take a mass of $10^{66}MeV$ which is of the same order of magnitude than the supermassive black hole in our galaxy, we obtain the same rotation curve but $r_{NF}$ would be displaced to $1.467.10^7$ kpc in disagreement with the almost flat part in the rotation curve of the Milky Way extending from 6 to about 16 kpc.

\section{FINAL COMMENTS}

Although this  study,  extended to the galactic scale, do not go beyond an initial exploratory stage, it seems valid to explain the rotation curves of low mass galactic nuclei, typically up to about  $1 M_\odot$.

Apart of the cosmic censorship hypothesis, we note that from the general relativity standpoint a simple electron, or positron, is a naked singularity. On the other hand, there is no evidence of a charged universe either locally or at the cluster or supercluster scale. We have seen that the obtained results do not depend on the sign of the charge. Hence, on a sufficient large scale (cluster or supercluster), at least in principle, no conflict with an a priory neutral universe might arise (for the hypothetical case of an overall charged Universe see \cite{overall}). How a charged naked singularity could be formed, possibly in the very early universe, is an open question.

Regarding the Dark Matter (DM) issue (proposed for the first time just a hundred years ago \cite{first}), since DM do not interact with the electromagnetic field, thereby its name, it is clear that the present study, like MOND, eludes the necessity of DM for the explanation of galactic rotation curves. 

At the time of writing, there is not any established theory of DM. Once such a theory is found, perhaps, it might even be compatible with the alternative proposed in this work where, according to general relativity, if charged collapsed objects exist, this describes the rotational curves at least in some range of masses irrespective of the existence of DM in galaxies.

The existence of dark matter is usually associated with all sorts of gauge fields and the local gauge invariance principle there being myriads of articles dealing with it. Unlike MOND, the present work is susceptible to being related to a U(1) local gauge symmetry or a Stueckelberg mechanism in the case of a massive field \cite{stuka} (this could be the subject of another work. 
\footnote{The connection between DM and U(1) have been the subject of research through many different ways including even String Theory \cite{string}. For some works following other lines of development, see, for instance, \cite{scipost} \cite{dudas} \cite{sm}}.

\section{CONFLICTS OF INTEREST}

The author declares that there are no conflicts of interest regarding the publication of this paper.
\end{document}